\documentclass{aa}  

\usepackage{graphicx}
\usepackage{txfonts}
\usepackage{multirow}
\usepackage{caption, threeparttable}
\begin{document}

\newcommand{\teff}{$T_\mathrm{eff}$}
\newcommand{\logg}{$\log g$}
\newcommand{\feh}{[Fe/H]}
\newcommand{\microturb}{$\xi_\mathrm{micro}$}
\newcommand{\sife}{[Si/Fe]}
\newcommand{\mgfe}{[Mg/Fe]}
\newcommand{\afe}{[$\alpha$/Fe]}
\newcommand{\co}{CO($\nu =2-0$)\,}

\newcommand{\water}{H$_2$O}
\newcommand{\invcm}{cm$^{-1}$}
\newcommand{\kms}{km\,s$^{-1}$}
\newcommand{\mic}{$\mu \mathrm m$}


   \title{M Giants with IGRINS\\
   II. Chemical Evolution of Fluorine at High Metallicities
}

   \subtitle{}

   \author{G. Nandakumar
            \inst{1}
            \and
          N. Ryde
          \inst{1}
            \and
        G. Mace
        \inst{2}
          }

   \institute{Lund Observatory, Division of Astrophysics, Department of Physics, Lund University, Box 43, SE-221 00 Lund, Sweden\\
              \email{govind.nandakumar@astro.lu.se}
              \and
    Department of Astronomy and McDonald Observatory, The University of Texas, Austin, TX 78712, USA}

   \date{Received ; accepted }

 
  \abstract
   { The origin and evolution of fluorine in the Milky Way galaxy is still in debate. 
   In particular, the increase of the [F/Fe] in metal-rich stars found from near-IR HF-lines is challenging to explain theoretically.   
   Chemical evolution models with current knowledge of yields from different fluorine producing stellar sources can not reproduce these observations. 
   }
   {The aim with this work is to observationally study the the Galactic chemical evolution of fluorine, especially for metal-rich stars. We want to investigate whether or not the significant rise of fluorine production at high metallicities can be corroborated. Furthermore, we want to explore possible reasons for this upturn in [F/Fe].
   }  
   {We determine the fluorine abundances from 50 M giants (3300<\teff<3800 K) in the  solar neighbourhood spanning a broad range of metallicities (-0.9 $<$ [Fe/H] $<$ 0.25 dex). These stars are cool enough to have an array of lines from the HF molecule in the K band. We observed the stars with the Immersion GRating INfrared Spectrograph (IGRINS) spectrometer   
   mounted on the Gemini South telescope and 
on the Harlan J. Smith Telescope at McDonald Observatory and investigate each of ten HF molecular lines in detail. }
   { Based on a detailed line-by-line analysis of ten HF lines, we 
   find that the R19, R18, and R16 lines (22699.49, 22714.59, and 22778.25 \AA) should primarily be used for an abundance analysis. The R15, R14, and R13 lines at 22826.86, 22886.73, and 22957.94 \AA\ can also be used, but the trends based on these lines show increasing dependencies with the stellar parameters. The strongest HF lines, namely R12, R11, R9, and R7 lying at 23040.57, 23134.76, 23358.33, and 23629.99 \AA\ should be avoided. The abundances derived from these strongest lines show significant trends with the stellar parameters, as well as a high sensitivity to variations in the stellar microturbulence, especially for coolest and most
metal-rich stars. This leads to a huge scatter and high fluorine abundances for supersolar metallicity stars, not seen in the trends from the weaker lines for the same stars. 
 }
   {When estimating the final mean fluorine abundance trend as a function metallicity, we neglect the fluorine abundances from the four strongest lines (R7, R9, R11, and R12) for all stars and use only those derived from R16, R18, and R19 for the coolest and most metal-rich stars. We confirm the flat trend of [F/Fe] found in other studies  for stars in the metallicity range of -1.0 $<$ \feh\,$<$ 0.0. We also find a slight enhancement at supersolar metallicities (0<\feh<0.15) but we can not confirm the upward trend seen at \feh\,$>$ 0.25. The HF line is intrinsically temperature sensitive which calls for studies of stars with highly accurate and homogeneous stellar parameters. The spread in our trend is presumably caused by the temperature sensitivity.    
We need more observations of M giants at super solar metallicities with a spectrometer that covers as many of the HF lines as possible, for instance the IGRINS spectrometer, to confirm if the metal-rich fluorine abundance upturn is real or not. }


   \maketitle
%

\section{Introduction}
\label{sec:intro}

 The chemical evolution of the Milky Way galaxy can be deciphered from the chemical nature of the stars belonging to the stellar populations that make up its components (Galactic disc, bulge, halo etc). 
The study of chemical evolution is enabled by the fact that the photospheric abundances of individual elements derived from absorption lines in stellar spectra trace the chemistry of the gas in the interstellar medium from which the stars were formed. The chemical evolution trends of observed elemental abundance ratios for different stellar populations can also open up the possibility to explore the dominant production sites or progenitors of each element by comparison with trends from theoretical chemical evolution models. 

While the elements like O, Mg, Si, P and Ti have been inferred to form in massive stars based on such investigations \citep{Chieffi:2004,cescutti:2012,Nomoto:2013,Matteucci:2021}, there are many elements whose origin and progenitors are still in debate. Fluorine is one such element. From theory, multiple production sites have been suggested, such as rapidly rotating, massive stars \citep{prantzos:18}, thermal pulses in Asymptotic Giant Branch (AGB) stars (\citealt{Forestini:1992}, \citealt{Straniero:2006}), the neutrino-process in core collapse supernovae \citep{Woosley:1988} as well as in novae \citep{Jose:1998,Spitoni:2018}. Whether or not helium burning phases in Wolf-Rayet (WR) stars \citep{Meynet:2000,palacios:2005} contribute to the cosmic budget of fluorine is uncertain.
From observations, multiple production sites have also been suggested, see for example \citet{Ryde:2020,Ryde:20:jaa}.
 
 A theoretical difficulty is the large uncertainties in the $^{19}$F yields (in M$_{\odot}$) from several of the above mentioned production sites like WR, novae etc, as well as differences in the mass and metallicity ranges adopted in different stellar yield calculations. There is also no final consensus on the initial rotational velocities adopted for massive stars and their metallicity dependencies. Thus, it is difficult to use chemical evolution models, that adopt different stellar yields as their input parameter, to pinpoint the exact production sites in different metallicity ranges. Many recent studies have explored the variation in the fluorine abundance trends resulting from chemical evolution models that adopt different combinations of the above mentioned parameters and assuming different progenitors \citep{Spitoni:2018,Grisoni:2020,Womack:2023}. 
 
 From an observational perspective as well, there are many challenges in the measurement of the fluorine abundances from stellar spectra. Firstly, the cosmic abundance of fluorine is generally very low. It is three orders of magnitude smaller than that of its immediate neighbors in the periodic table (C, N, O, Ne, Na, Mg, Al, and Si). This is a reflection of its unique formation channels, where Galactic fluorine is not synthesised in the main nuclear burning phases of stars \citep{clayton:2003}. In spite of the fact that newly formed fluorine nuclei in stellar interiors readily react with hydrogen and helium, there are, however, processes that can synthesize it in order for it to survive and contribute to the buildup of the cosmic reservoir of fluorine. Secondly, there are not many diagnostically useful spectra lines from which the fluorine abundances can be determined. There are no strong atomic fluorine lines at optical or infrared wavelengths and there are only a few accessible molecular lines in the form of vibration-rotational line from the HF molecules in the rather crowded as well as telluric affected infrared regimes.

 The HF lines have been used to determine fluorine abundances in AGB stars \citep{Jorissen:1992,Abia:2009,Abia:2010,Abia:2015,Abia:2019}, giants in globular clusters \citep{deLaverny:2013a,deLaverny:2013b,Guerco:2019a}, in dwarfs \citep{Recio-Blanco:2012}, field G, K and M giants in the Galactic disc \citep{jonsson:2014a,Pilachowski:2015,jonsson:2017,Guerco:2019b,Ryde:2020,Ryde:20:jaa}, Galactic bulge stars \citep{jonsson:2014b}, and recently in the Galactic nuclear star cluster \citep{Guerco:2022}.  Among these studies, only a handful have determined the fluorine abundances for stars at super-solar metallicities, i.e. stars with \feh>0, see for example \citet{jonsson:2017,Ryde:2020,Guerco:2022}. The general finding in these studies is an enhanced fluorine-to-iron abundance ratio, as well as a significant upturn with increasing metallicity, which is interpreted as a secondary behaviour of fluorine \citep{Ryde:2020}. Chemical evolution models have been unable to reproduce this trend for thin and thick disk stars \citep[e.g.][]{Spitoni:2018}. Suggested reasons for this include large uncertainties concerning the nucleosynthesis of fluorine \citep{Spitoni:2018}, the need for invoking uncertain production sites like AGB stars contributing at later times, high mass-loss from metal-rich, massive stars, and/or 
 novae \citep{Grisoni:2020}, large uncertainties in the fluorine yields at super-solar metallicities, as well as differences in the stellar composition compared to the local gas of the interstellar medium. The latter would be expected since at super-solar metallicities, stars are expected to have formed in the inner disk and migrated to the solar neighborhood \citep{Womack:2023}. Thus, this upturn of fluorine at super-solar metallicities still remains a mystery, and is the subject of this paper.

 The vibrational-rotational molecular line of HF at $\lambda_\mathrm{air}=23358.33$\,\AA\ is most commonly used to determine fluorine abundance in the above mentioned studies. This is primarily owing to the fact that the  line is sufficiently strong, least affected (blended) by lines of other elements or molecules contained in the stellar atmosphere, and not affected by telluric lines. Other HF lines get considerably weaker for spectral types hotter than M type (\teff\,$>$ 3900 K), which are the stellar types used in most previous studies. Another reason is that some  instruments used to obtain K-band spectra do not record the full K band and will have  gaps in wavelength. This might result in the absence of many of the other HF lines in the observed spectra. 
 The first issue can be solved by targeting and obtaining the spectra of the cooler giants of M type (\teff\,$<$ 3900 K), which will have stronger HF lines. An instrument such as the Immersion GRating INfrared Spectrograph \citep[IGRINS;][]{Yuk:2010,Wang:2010,Gully:2012,Moon:2012,Park:2014,Jeong:2014} provides spectra spanning the full H and K bands (1.45 - 2.5 $\mu$m), which therefore will be able to record all possible R-branch lines of HF, all lying in the K band, solving the second issue. 
 
 In this paper, we use IGRINS spectra of 50  cool giants in the solar neighbourhood, stars of spectral type M spanning a broad range of metallicities ($-0.9 <$ \feh\, $< 0.25$\, dex).  For these stars we determine the fluorine abundances from ten HF molecular lines. We carry out detailed line-by-line analysis and plot individual fluorine abundance trends as a function of \teff, \logg, \feh, and $\xi_\mathrm{micro}$ to investigate whether a similar upturn in fluorine abundances at super-solar metallicities is evident for the cool M giants as well.
 
Our observations and data reduction procedure are described in section~\ref{sec:obs}. The determination of the fluorine abundance 
using the Spectroscopy Made Easy (SME) code is described in Section~\ref{sec:analysis}, followed by results and discussion in Sections~\ref{sec:results} and \ref{sec:discussion}, respectively. Finally, we make some concluding remarks in Section~\ref{sec:conclusion}.
 

\section{Observations and Data reduction}
\label{sec:obs}
We thus determine the fluorine abundance for 50 late-K to M stars from high-resolution K-band spectra observed with the IGRINS spectrograph. 
IGRINS provides a spectral resolving power of $R \sim$ 45,000, and the reductions were done with the IGRINS PipeLine Package \citep[IGRINS PLP;][]{Lee:2017} to optimally extract the telluric corrected, wavelength calibrated spectra after flat-field correction \citep{Han:2012,Oh:2014}. The spectra were then resampled and normalized in {\tt iraf} \citep{IRAF} but to take care of any residual modulations in the continuum levels, we put large attention in defining specific local continua around the HF line being studied. Finally, the spectra are shifted to laboratory wavelengths in air after a stellar radial velocity correction. The average signal-to-noise ratios (SNR)\footnote{SNR is provided by RRISA \citep[The Raw $\&$ Reduced IGRINS Spectral Archive;][]{rrisa} and is the average SNR for K band and is per resolution element. It varies over the orders and it is lowest at the ends of the orders} of the spectra of our stars range from  65-400, with most spectra having SNR$>100$ \citet{Nandakumar:2023}.

The 50 stars were also analysed for their fundamental parameters in \citet{Nandakumar:2023}, where further details of the observations and data reduction are given. Apart from 6 stars from the IGRINS spectral library archive \citep{Park:2018,rrisa}, which were observed at McDonald Observatory, the stars were observed at the Gemini South telescope \citep{Mace:2018} within the programs GS-2020B-Q-305 and GS-2021A-Q302 in Jan to April 2021. 

\begin{table}
\caption{HF line data for the R branch ($\nu'' = 0$ to $\nu' = 1$) lines used in this study. }\label{table:lines}
\begin{tabular}{c c c c c }
\hline
\hline
 Molecule &  & $\lambda_{Air}$  & $\chi$ & log $gf$   \\
\hline
&  & \AA & eV &  \\
 \hline
  H$^{19}$F & R7 & 23629.991 & 0.142 & -4.004 \\
  H$^{19}$F & R9 & 23358.329 & 0.227 & -3.962 \\
  H$^{19}$F & R11 & 23134.757 & 0.332 & -3.942 \\
  H$^{19}$F & R12 & 23040.574 & 0.391 & -3.940 \\
  H$^{19}$F & R13 & 22957.938 & 0.455 & -3.941 \\
  H$^{19}$F & R14 & 22886.733 & 0.524 & -3.947 \\
  H$^{19}$F & R15 & 22826.862 & 0.597 & -3.956 \\
  H$^{19}$F & R16 & 22778.249 & 0.674 & -3.969 \\
  H$^{19}$F & R18 & 22714.589 & 0.842 & -4.007 \\
  H$^{19}$F & R19 & 22699.488 & 0.932 & -4.031 \\
\hline
\hline
\end{tabular}
 
\end{table}

\begin{table*}
\caption{Stellar parameters, [C/Fe], [N/Fe], [O/Fe] and mean [F/Fe] values along with the standard error of mean (from line-by-line abundances) for each star determined in this work. }\label{table:parameters}
\begin{tabular}{c c c c c c c c c c c}
\hline
\hline
 Index & Star  & T$_\mathrm{eff}$ & $\log g$  & [Fe/H]  &  $\xi_\mathrm{micro}$ & [C/Fe] & [N/Fe] & [O/Fe]  & $<$[F/Fe]$>$ & $\sigma$[F/Fe]  \\
 \hline
  & & K & log(cm/s$^{2}$) & dex & Km/s & dex & dex & dex & dex & dex \\
  \hline
1  &  2M05484106-0602007  &  3490  &  0.48  &  -0.28  &  2.03  &  -0.02  &  0.17  &  0.1  &  -0.17  &  0.02  \\  
2  &  2M05594446-7212111  &  3694  &  0.74  &  -0.45  &  1.88  &  0.02  &  0.19  &  0.16  &  -0.12  &  0.0  \\  
3  &  2M06035110-7456029  &  3562  &  0.48  &  -0.51  &  2.14  &  0.21  &  0.3  &  0.38  &  -0.02  &  0.04  \\  
4  &  2M06035214-7255079  &  3742  &  1.08  &  0.0  &  1.78  &  0.08  &  0.29  &  0.15  &  -0.1  &  0.02  \\  
5  &  2M06052796-0553384  &  3677  &  0.92  &  -0.07  &  1.78  &  -0.12  &  0.21  &  0.04  &  -0.25  &  0.03  \\  
6  &  2M06074096-0530332  &  3692  &  0.68  &  -0.54  &  2.01  &  0.02  &  0.31  &  0.19  &  -0.07  &  0.01  \\  
7  &  2M06124201-0025095  &  3583  &  0.42  &  -0.66  &  2.39  &  0.05  &  0.33  &  0.24  &  -0.04  &  0.0  \\  
8  &  2M06140107-0641072  &  3620  &  0.65  &  -0.38  &  2.1  &  0.17  &  0.22  &  0.33  &  -0.12  &  0.02  \\  
9  &  2M06143705-0551064  &  3608  &  0.55  &  -0.52  &  2.24  &  0.06  &  0.27  &  0.19  &  -0.1  &  0.03  \\  
10  &  2M06171159-7259319  &  3800  &  0.59  &  -0.92  &  2.72  &  0.21  &  0.37  &  0.56  &  -0.08  &  0.04  \\ 
11  &  2M06223443-0443153  &  3521  &  0.4  &  -0.52  &  2.19  &  0.0  &  0.31  &  0.18  &  -0.07  &  0.04  \\  
12  &  2M06231693-0530385  &  3484  &  0.32  &  -0.55  &  2.09  &  0.04  &  0.3  &  0.2  &  -0.08  &  0.02  \\  
13  &  2M06520463-0047080  &  3581  &  0.67  &  -0.21  &  2.15  &  -0.09  &  0.31  &  0.08  &  -0.13  &  0.03  \\  
14  &  2M06551808-0148080  &  3606  &  0.52  &  -0.56  &  1.96  &  0.04  &  0.27  &  0.2  &  -0.15  &  0.01  \\  
15  &  2M06574070-1231239  &  3561  &  0.56  &  -0.36  &  2.24  &  -0.08  &  0.41  &  0.13  &  -0.06  &  0.01  \\  
16  &  2M10430394-4605354  &  3568  &  0.96  &  0.25  &  1.83  &  -0.18  &  0.39  &  -0.08  &  -0.09  &  0.03  \\  
17  &  2M11042542-7318068  &  3566  &  0.51  &  -0.46  &  2.02  &  0.01  &  0.24  &  0.16  &  -0.12  &  0.0  \\  
18  &  2M12101600-4936072  &  3539  &  0.5  &  -0.41  &  2.05  &  0.03  &  0.23  &  0.15  &  -0.11  &  0.03  \\  
19  &  2M13403516-5040261  &  3528  &  0.61  &  -0.15  &  1.92  &  -0.02  &  0.13  &  0.06  &  -0.1  &  0.03  \\  
20  &  2M14131192-4849280  &  3504  &  0.61  &  -0.08  &  1.81  &  0.01  &  0.16  &  0.03  &  -0.03  &  0.02  \\  
21  &  2M14240039-6252516  &  3474  &  0.69  &  0.12  &  1.94  &  -0.07  &  0.21  &  -0.04  &  0.04  &  0.04  \\  
22  &  2M14241044-6218367  &  3543  &  0.8  &  0.11  &  1.95  &  -0.05  &  0.2  &  -0.04  &  -0.03  &  0.02  \\  
23  &  2M14260433-6219024  &  3386  &  0.55  &  0.13  &  1.82  &  -0.04  &  0.14  &  -0.05  &  -0.06  &  0.02  \\  
24  &  2M14261117-6240220  &  3387  &  0.52  &  0.08  &  1.92  &  -0.03  &  0.21  &  -0.03  &  0.05  &  0.01  \\  
25  &  2M14275833-6147534  &  3453  &  0.63  &  0.08  &  1.91  &  -0.1  &  0.22  &  -0.03  &  -0.01  &  0.02  \\  
26  &  2M14283733-6257279  &  3465  &  0.62  &  0.04  &  1.83  &  -0.09  &  0.33  &  -0.02  &  0.01  &  0.02  \\  
27  &  2M14291063-6317181  &  3430  &  0.54  &  0.0  &  1.95  &  -0.02  &  0.16  &  -0.0  &  -0.03  &  0.03  \\  
28  &  2M14311520-6145468  &  3499  &  0.62  &  -0.06  &  2.01  &  -0.07  &  0.25  &  0.02  &  -0.02  &  0.02  \\  
29  &  2M14322072-6215506  &  3639  &  0.89  &  -0.0  &  1.76  &  -0.05  &  0.08  &  0.01  &  -0.27  &  0.03  \\  
30  &  2M14332169-6302108  &  3524  &  0.56  &  -0.25  &  1.98  &  0.02  &  0.15  &  0.09  &  -0.12  &  0.04  \\  
31  &  2M14332869-6211255  &  3664  &  1.11  &  0.23  &  1.99  &  -0.09  &  0.35  &  -0.08  &  -0.09  &  0.0  \\  
32  &  2M14333081-6221450  &  3430  &  0.55  &  0.02  &  1.92  &  -0.09  &  0.21  &  -0.01  &  -0.03  &  0.02  \\  
33  &  2M14333688-6232028  &  3425  &  0.54  &  0.02  &  1.87  &  -0.07  &  0.25  &  -0.01  &  -0.02  &  0.03  \\  
34  &  2M14345114-6225509  &  3442  &  0.68  &  0.18  &  1.85  &  -0.0  &  0.16  &  -0.06  &  0.03  &  0.02  \\  
35  &  2M14360142-6228561  &  3514  &  0.53  &  -0.26  &  2.02  &  -0.04  &  0.16  &  0.09  &  -0.09  &  0.02  \\  
36  &  2M14360935-6309399  &  3446  &  0.61  &  0.08  &  1.99  &  -0.04  &  0.21  &  -0.03  &  0.13  &  0.01  \\  
37  &  2M14371958-6251344  &  3650  &  0.98  &  0.1  &  1.8  &  -0.11  &  0.29  &  -0.04  &  -0.12  &  0.02  \\  
38  &  2M14375085-6237526  &  3582  &  0.96  &  0.23  &  1.8  &  -0.11  &  0.31  &  -0.09  &  -0.02  &  0.01  \\  
39  &  2M15161949+0244516  &  3691  &  0.76  &  -0.4  &  1.98  &  0.17  &  0.14  &  0.34  &  -0.15  &  0.01  \\  
40  &  2M17584888-2351011  &  3564  &  0.95  &  0.25  &  2.2  &  -0.04  &  0.38  &  -0.09  &  -0.03  &  0.02  \\  
41  &  2M18103303-1626220  &  3347  &  0.46  &  0.09  &  1.98  &  -0.09  &  0.21  &  -0.03  &  0.09  &  0.02  \\  
42  &  2M18142346-2136410  &  3390  &  0.48  &  0.01  &  1.96  &  -0.01  &  0.27  &  -0.0  &  0.16  &  0.02  \\  
43  &  2M18191551-1726223  &  3434  &  0.59  &  0.07  &  1.93  &  -0.07  &  0.24  &  -0.02  &  0.02  &  0.03  \\  
44  &  2M18522108-3022143  &  3578  &  0.45  &  -0.59  &  2.26  &  0.08  &  0.54  &  0.41  &  -0.23  &  0.02  \\  
45  &  HD132813  &  3457  &  0.44  &  -0.27  &  1.88  &  0.0  &  0.14  &  0.1  &  0.06  &  0.01  \\  
46  &  HD175588  &  3484  &  0.6  &  -0.04  &  2.24  &  -0.02  &  0.34  &  0.02  &  0.03  &  0.04  \\  
47  &  HD89758  &  3807  &  1.15  &  -0.09  &  1.65  &  -0.09  &  0.21  &  0.03  &  -0.03  &  0.06  \\  
48  &  HD224935  &  3529  &  0.64  &  -0.1  &  2.01  &  -0.08  &  0.17  &  0.04  &  -0.02  &  0.02  \\  
49  &  HD101153  &  3438  &  0.51  &  -0.07  &  2.03  &  -0.07  &  0.2  &  0.03  &  0.05  &  0.03  \\  
50  &  HIP54396  &  3459  &  0.5  &  -0.15  &  1.86  &  0.04  &  0.07  &  0.05  &  -0.02  &  0.06  \\  
\hline

\hline
\hline
\end{tabular}
 
\end{table*}

 \begin{figure*}
  \includegraphics[width=\textwidth]{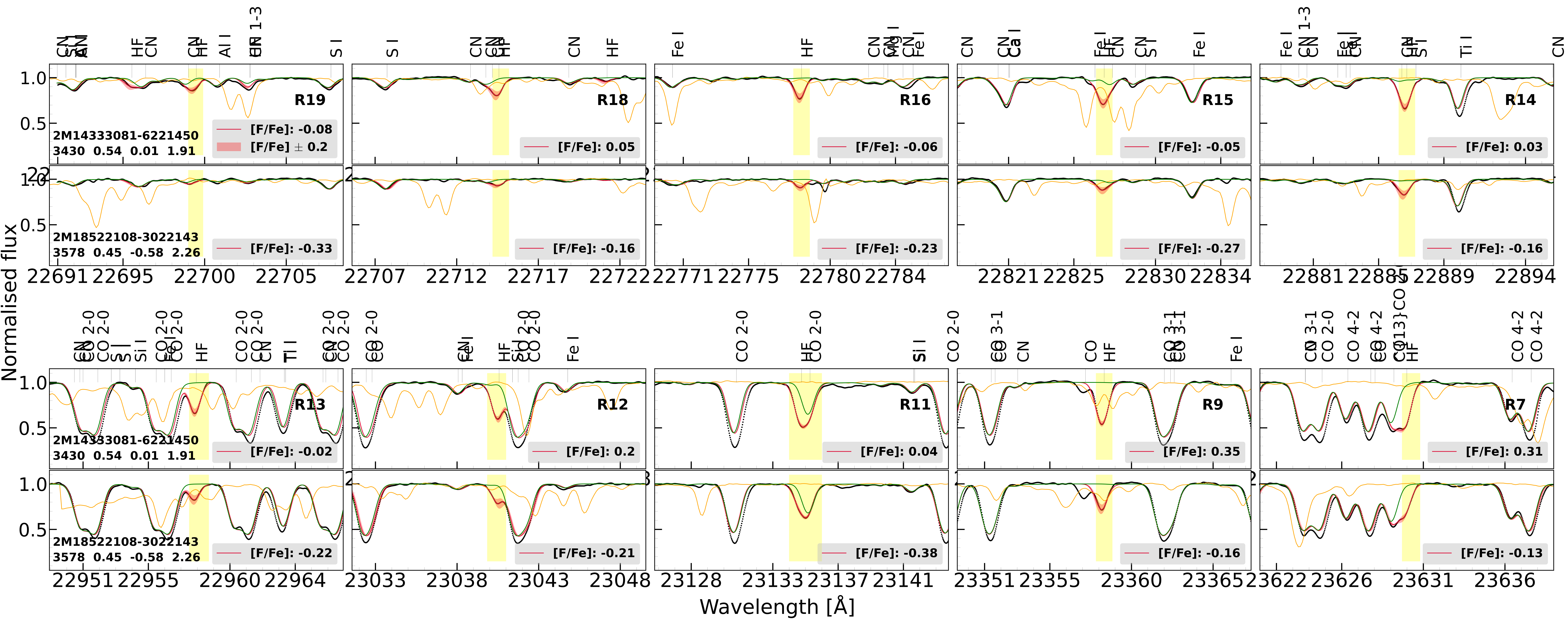}
  \caption{ The ten HF lines used 
  to determine the fluorine abundances.  The top two row panels 
  show the observed spectra (black circles) for 5 HF lines (R19, R18, R16, R15, and R14) of one 
  metal rich star 
  (32) and one metal poor star 
  (44), with the stellar parameters (\teff, \logg, [Fe/H], and \microturb) given to the left. The bottom two row panels show the observed spectra for the remaining 5 HF lines (R13, R12, R11, R9, and R7) of the same two stars. 
  The orange line denote where the telluric lines lie before the telluric correction,  and the red band the variation in the synthetic spectrum for $\pm$ 0.2 dex difference in fluorine abundance. 
  The green line shows the synthetic spectrum without HF. 
  The [F/Fe] values 
  are also listed for each line. All identified atomic and molecular lines are  denoted above the spectra. } 
  \label{fig:flourine_spectra}%
\end{figure*}

\begin{table*}
\caption{ [F/Fe] determined from the six HF lines after careful visual inspection of every star spectrum. The fluorine abundances from problematic lines (noisy or affected by telluric residuals) are indicated by "--".  }\label{table:Fabund}
\centering
\begin{tabular}{c c c c c c c c c c c}
\hline
\hline
 Index & R19  & R18  &  R16  &  R15  &  R14  &  R13  &  R12  &  R11  &  R9  &  R7     \\
\hline
1  &  -0.19  &  -0.07  &  -0.26  &  -0.22  &  --  &  -0.15  &  -0.15  &  -0.28  &  -0.08  &  0.0  \\ 
2  &  --  &  --  &  --  &  --  &  -0.12  &  --  &  --  &  -0.23  &  -0.14  &  --    \\ 
3  &  -0.02  &  --  &  -0.09  &  -0.09  &  0.14  &  --  &  --  &  -0.11  &  0.08  &  --    \\ 
4  &  --  &  --  &  -0.14  &  -0.09  &  -0.15  &  -0.04  &  --  &  -0.12  &  -0.03  &  --    \\ 
5  &  -0.17  &  --  &  --  &  -0.34  &  -0.19  &  -0.32  &  -0.24  &  -0.35  &  -0.2  &  --    \\ 
6  &  --  &  --  &  -0.09  &  -0.09  &  -0.05  &  -0.04  &  -0.08  &  -0.13  &  -0.06  &  --    \\ 
7  &  --  &  --  &  -0.04  &  -0.05  &  -0.04  &  --  &  --  &  -0.2  &  0.07  &  --    \\ 
8  &  --  &  --  &  -0.13  &  -0.16  &  -0.07  &  --  &  --  &  -0.15  &  -0.12  &  --    \\ 
9  &  --  &  --  &  -0.13  &  -0.05  &  -0.01  &  -0.19  &  --  &  -0.14  &  -0.04  &  --    \\ 
10  &  --  &  --  &  --  &  -0.16  &  0.01  &  --  &  --  &  -0.41  &  -0.23  &  --    \\ 
11  &  -0.14  &  0.09  &  -0.09  &  --  &  --  &  -0.14  &  --  &  -0.19  &  0.01  &  --    \\ 
12  &  -0.05  &  -0.07  &  -0.09  &  -0.13  &  -0.07  &  -0.18  &  0.04  &  -0.3  &  0.01  &  -0.15  \\ 
13  &  -0.07  &  --  &  -0.16  &  -0.15  &  --  &  -0.21  &  -0.05  &  -0.12  &  -0.04  &  -0.08  \\ 
14  &  -0.2  &  --  &  -0.15  &  -0.18  &  -0.06  &  -0.17  &  -0.13  &  -0.23  &  -0.09  &  -0.09  \\ 
15  &  --  &  --  &  -0.07  &  --  &  --  &  -0.06  &  -0.04  &  -0.1  &  -0.0  &  -0.03  \\ 
16  &  --  &  --  &  -0.22  &  -0.13  &  -0.06  &  -0.07  &  0.03  &  -0.21  &  0.06  &  0.07  \\ 
17  &  --  &  --  &  --  &  --  &  -0.12  &  --  &  --  &  -0.14  &  -0.06  &  --    \\ 
18  &  --  &  --  &  -0.14  &  -0.14  &  0.03  &  -0.2  &  --  &  -0.12  &  0.02  &  0.08  \\ 
19  &  --  &  -0.03  &  -0.17  &  --  &  -0.12  &  -0.16  &  -0.02  &  -0.13  &  -0.02  &  0.12  \\ 
20  &  --  &  0.03  &  -0.06  &  --  &  -0.06  &  -0.04  &  --  &  -0.05  &  0.13  &  0.25  \\ 
21  &  --  &  0.12  &  -0.04  &  --  &  0.05  &  0.09  &  --  &  0.08  &  0.24  &  0.36  \\ 
22  &  --  &  0.03  &  -0.07  &  -0.12  &  -0.03  &  -0.04  &  0.05  &  -0.05  &  0.1  &  0.12  \\ 
23  &  0.0  &  -0.1  &  -0.07  &  --  &  0.06  &  0.05  &  --  &  0.03  &  0.29  &  --    \\ 
24  &  --  &  0.08  &  0.03  &  0.01  &  0.08  &  -0.01  &  0.2  &  0.21  &  0.34  &  0.31  \\ 
25  &  -0.07  &  0.06  &  -0.01  &  --  &  0.06  &  0.1  &  0.26  &  0.13  &  0.21  &  0.25  \\ 
26  &  --  &  0.04  &  -0.03  &  --  &  -0.05  &  -0.01  &  --  &  0.01  &  0.18  &  --    \\ 
27  &  -0.06  &  0.08  &  -0.1  &  -0.01  &  -0.05  &  --  &  0.18  &  0.06  &  0.19  &  0.25  \\ 
28  &  -0.11  &  0.13  &  -0.05  &  --  &  -0.05  &  -0.06  &  0.03  &  -0.03  &  0.08  &  --    \\ 
29  &  --  &  -0.13  &  --  &  --  &  -0.33  &  -0.33  &  -0.27  &  -0.48  &  -0.28  &  -0.2  \\ 
30  &  --  &  -0.02  &  -0.22  &  --  &  -0.11  &  --  &  --  &  -0.2  &  -0.04  &  --    \\ 
31  &  --  &  -0.1  &  -0.1  &  --  &  -0.09  &  --  &  --  &  -0.2  &  0.01  &  0.04  \\ 
32  &  -0.08  &  0.05  &  -0.06  &  -0.07  &  0.03  &  -0.02  &  0.2  &  0.04  &  0.36  &  0.31  \\ 
33  &  -0.12  &  0.05  &  0.01  &  0.01  &  0.12  &  0.06  &  0.28  &  0.18  &  0.3  &  0.48  \\ 
34  &  -0.02  &  0.07  &  0.05  &  0.05  &  0.08  &  0.12  &  --  &  0.24  &  0.43  &  --    \\ 
35  &  --  &  -0.04  &  -0.09  &  -0.15  &  -0.08  &  --  &  --  &  -0.15  &  0.0  &  -0.03  \\ 
36  &  0.12  &  0.17  &  0.11  &  --  &  0.15  &  --  &  0.48  &  0.31  &  0.48  &  0.73  \\ 
37  &  -0.12  &  -0.01  &  -0.18  &  -0.19  &  --  &  -0.11  &  -0.08  &  -0.19  &  -0.03  &  -0.04  \\ 
38  &  --  &  -0.03  &  -0.02  &  --  &  -0.01  &  --  &  --  &  -0.14  &  0.21  &  --    \\ 
39  &  --  &  --  &  -0.17  &  --  &  -0.14  &  --  &  --  &  -0.35  &  -0.2  &  --    \\ 
40  &  -0.15  &  -0.04  &  -0.09  &  -0.13  &  -0.05  &  -0.07  &  0.32  &  -0.11  &  0.09  &  0.09  \\ 
41  &  0.03  &  0.14  &  0.11  &  0.09  &  0.29  &  0.25  &  --  &  0.28  &  0.56  &  0.55  \\ 
42  &  --  &  0.2  &  0.11  &  0.09  &  0.26  &  0.15  &  0.5  &  0.34  &  0.54  &  0.62  \\ 
43  &  -0.05  &  0.1  &  0.01  &  --  &  0.14  &  0.09  &  0.28  &  0.17  &  0.28  &  0.36  \\ 
44  &  -0.33  &  -0.16  &  -0.23  &  -0.27  &  -0.16  &  -0.22  &  -0.21  &  -0.38  &  -0.16  &  -0.13  \\ 
45  &  0.06  &  0.13  &  -0.01  &  0.04  &  0.08  &  0.07  &  0.3  &  0.31  &  0.33  &  0.47  \\ 
46  &  --  &  0.24  &  -0.05  &  -0.08  &  0.06  &  -0.01  &  0.32  &  0.26  &  0.31  &  0.36  \\ 
47  &  --  &  0.08  &  --  &  --  &  -0.15  &  --  &  --  &  -0.14  &  -0.12  &  -0.12  \\ 
48  &  -0.08  &  0.18  &  -0.08  &  -0.01  &  -0.06  &  -0.07  &  0.19  &  0.23  &  0.19  &  0.76  \\ 
49  &  -0.04  &  0.11  &  -0.01  &  --  &  0.14  &  0.07  &  0.22  &  0.35  &  0.36  &  0.49  \\ 
50  &  -0.21  &  --  &  -0.05  &  --  &  0.08  &  0.11  &  --  &  0.33  &  0.36  &  0.44  \\ 

\hline
\hline                                 
\end{tabular}

\end{table*}

\section{Analysis}
\label{sec:analysis}

The fluorine abundances are derived from vibrational-rotational lines from the HF molecule. For a given star, defined by its stellar parameters, \teff, \logg, \feh, and  \microturb\ \citep[determined by][]{Nandakumar:2023}, we synthesise model spectra using the Spectroscopy Made Easy code \citep[SME;][]{sme,sme_code}. The synthesis uses one-dimensional (1D) MARCS (Model Atmospheres in a Radiative and Convective Scheme) stellar atmosphere models \citep{marcs:08}, which are hydrostatic in spherical geometry, computed assuming LTE, chemical equilibrium, homogeneity, and conservation of the total flux. 
A relevant stellar atmosphere model is chosen by interpolating in a grid of MARCS models. The fluorine abundance is then set free and SME generates and fits multiple synthetic spectra with the fluorine abundance varying. The final stellar abundance derived from  the spectral line corresponds to the synthetic spectrum that best matches the observed spectrum by means of $\chi^{2}$ minimisation method.

In \citet{jonsson:2014b} a comprehensive line list of the  vibrational-rotational H$^{19}$F lines (the R and P branches of the $\nu'' = 0$ to $\nu' = 1$ band) in the K and L spectral bands are provided. The R branch lines lie toward the high wavelength edge of the K band, with the R4-R9\footnote{The R-number indicates the upper rotational level's quantum number, $J''$} lines being the strongest. The transition probabilities of the R7 to R19 lines do not vary much, but the series get increasingly weaker with increasing excitation energy of the lower rotational level, i.e. with increasing R-number. We have been able to derive abundances from 10 lines, namely the R7, R9, R11, R12, R13, R14, R15, R16, R18, and R19 lines. The other lines are heavily affected by blends or telluric lines and cannot be used to derive abundances. 

The details of the ten HF absorption lines
are listed in the Table~\ref{table:lines}. For the surrounding atomic lines we use the modified line-list described in \citet{Nandakumar:2023}. The line data for the CO, CN, and OH molecular lines are adopted from the line lists of \citet{li:2015}, \citet{brooke:2016} and \citet{sneden:2014}, respectively.

\subsection{Fundamental stellar parameters}
\label{sec:stellarparameters}

\citet{Nandakumar:2023} determined \teff, \logg, \feh, $\xi_\mathrm{micro}$, [C/Fe], and [N/Fe] simultaneously with an iterative spectroscopic method.  
In this method, the effective temperature is mainly constrained by selected \teff-sensitive OH lines, the metallicity by selected Fe lines, the microturbulence by different sets of weak and strong lines, and the C and N abundances by selected CO, CN molecular lines. These abundances ensures that CO and CN lines are fitted, which is especially important for the HF lines that are blended in part with CN and/or CO lines. The surface gravity is determined based on the effective temperature and metallicity by means of the Yonsei-Yale (YY) isochrones assuming old ages of 2-10 Gyr \citep{Demarque:2004}, which is appropriate for low-mass giants. The final stellar parameters and C, N, O abundances determined for the 50 solar neighbourhood stars are listed in the Table~\ref{table:parameters}.

 \begin{figure*}
  \includegraphics[width=\textwidth]{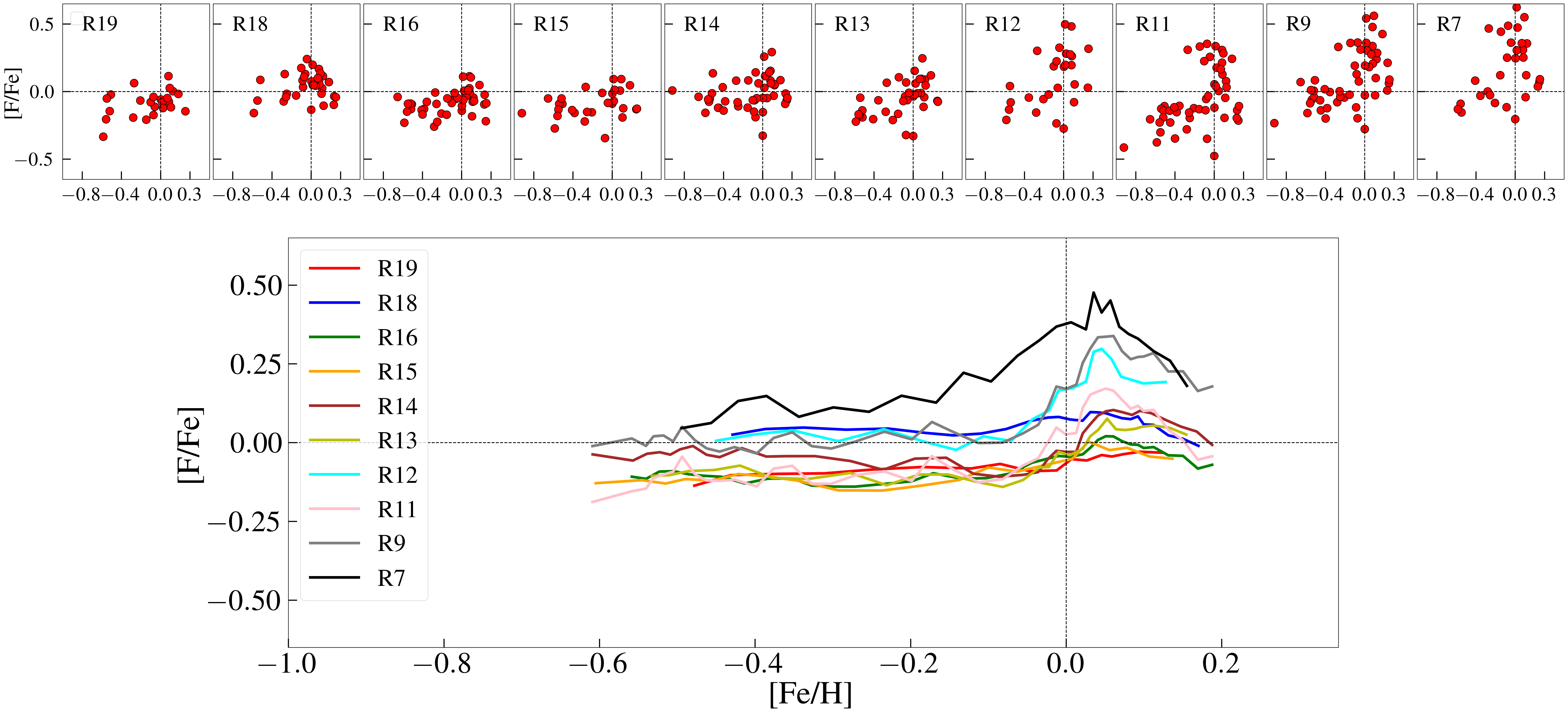}
  \caption{ Top panels: [F/Fe] versus [Fe/H] for the 50 solar neighborhood M giants (red circles) determined from the six K band molecular HF lines (each panel) in the IGRINS spectra. All are scaled to the following solar abundances: A(F)$_{\odot}$ = 4.43 \citep{Lodders:2003} and A(Fe)$_{\odot}$ = 7.45 \citep{solar:sme}. There is an evident increase and large scatter in [F/Fe] determined from the four reddest and strongest HF lines at 23040.57 \AA\, (R12), 23134.76 \AA\, (R11), 23358 \AA\, (R9), and 23629.99 \AA\, (R7) for stars with [Fe/H] $\sim$ 0 and above (not evident in other lines)}
  \label{fig:flourine_rm}%
\end{figure*}

\subsection{Determination of F abundance}
\label{sec:Fabund}

We determine individual fluorine abundances from each of the ten HF lines for all studied stars. In Figure~\ref{fig:flourine_spectra} we show the synthetic spectra fit (crimson line) to the first 5 HF lines (R19, R18, R16, R15, and R14; top two row panels) and the last 5 HF lines (R13, R12, R11, R9, and R7; bottom two row panels) in the resampled observed spectra (black circles) of one metal rich star 2M14333081-6221450 (32) and one metal poor star 2M18522108-3022143 (44). We demonstrate the sensitivity of each HF line to the fluorine abundance with the red band that shows the variation of the synthetic spectra with $\Delta$[F/Fe] = $\pm$0.2 dex. We also indicate the star-specific telluric lines used for telluric correction in orange to highlight where they lie and thus where some residuals might prevail as well as where the noise is expected to be larger than in the rest of the spectrum. Note, that the elimination of the telluric lines in general works very well. The yellow bands in each panel represent the line masks defined for the HF lines wherein SME fits the observed spectra by varying the fluorine abundance and finds the best synthetic spectra fit by chi-square minimisation. The green line shows the synthetic spectrum without HF, thus indicating any possible known blends lying in the wavelength range of the HF line. We note that the line R11 has a dominant CO blend, the R7 and R12 lines have very strong CO lines to the left and right respectively, and the R14, R15, R18, and R19 lines are blended by weak CN lines. The rest of the lines, namely, R16, R13, and R9 are the ones least affected by neighbouring lines or molecules. The [F/Fe] values corresponding to the best fit case for each line is also listed in each panel.

Further, we carried out a detailed visual inspection of each HF line in every stellar spectrum to select only the lines of highest quality, thus being unaffected by noise, spurious features, and/or bad telluric corrections. Based on this, we found the HF lines at 23134.76 \AA\ (R11), and 23358.33 \AA\, (R9) (the third and fourth lines in the bottom panels in Figure~\ref{fig:flourine_spectra}) 
to be of good quality for all 50 stars. The next best quality lines are HF lines at 22778.25 \AA\, (R16) and 22886.73 \AA\, (R14) that are found to be of low quality only for around four to six stars. The HF line at 22699.49 \AA\, (first line in the top panels shown in the Figure) is found to have the lowest quality, which makes it useful in only 
$\sim$ 50$\%$ of the stars. The finally selected fluorine abundances, [F/Fe], determined from each HF line are listed in the Table~\ref{table:Fabund}.

\begin{figure*}
  \includegraphics[width=\textwidth]{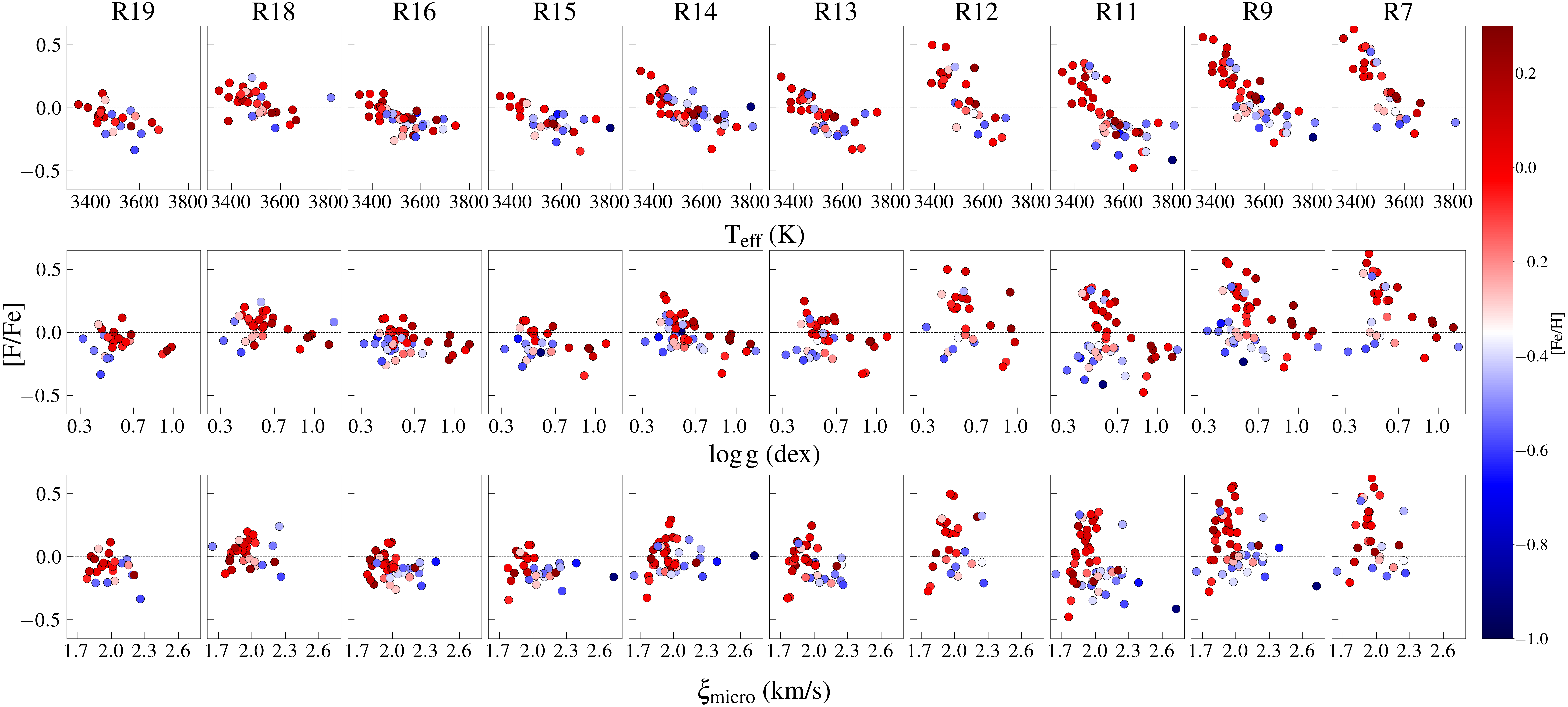}
  \caption{ [F/Fe] versus \teff\,(top panels), \logg\,(middle panels) and $\xi_\mathrm{micro}$ (bottom panels) color coded with respective metallicity with the horizontal panels in each row arranged in the increasing order of wavelengths of 10 HF lines from which [F/Fe] has been determined. }
  \label{fig:flourine_parameters}%
\end{figure*}

\begin{figure*}
  \includegraphics[width=\textwidth]{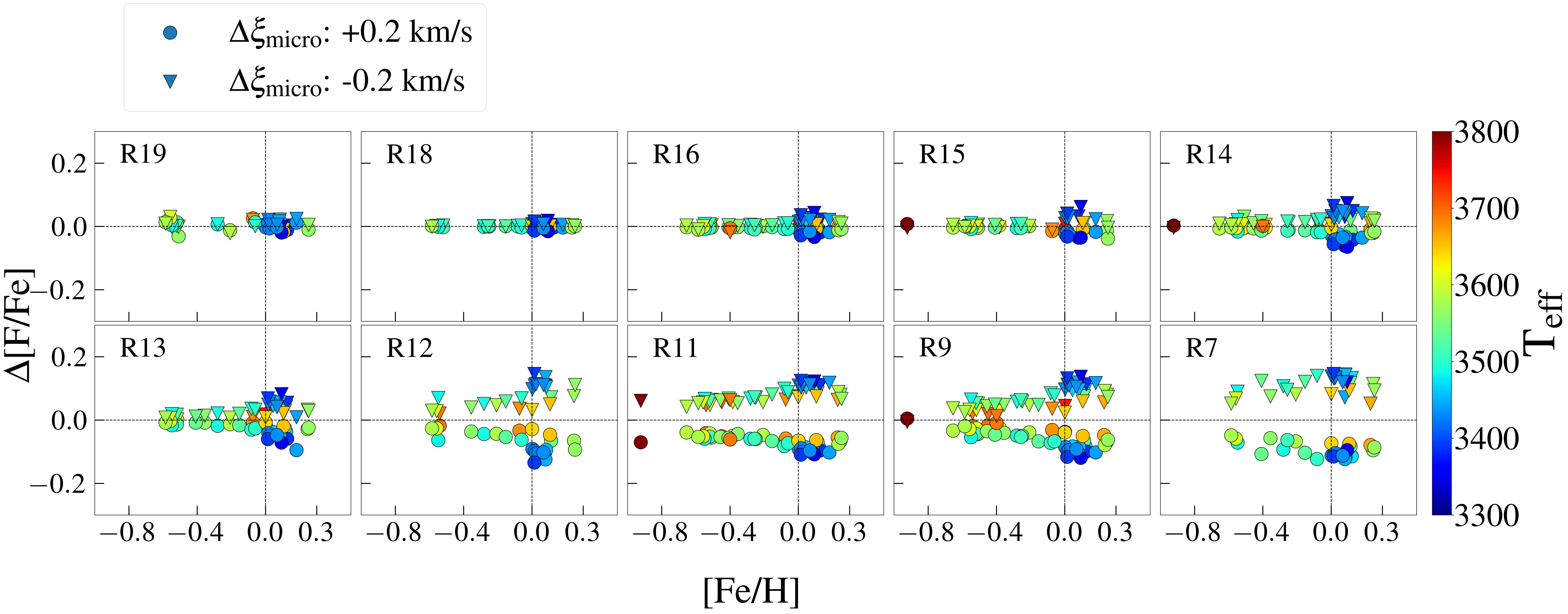}
  \caption{ Change in [F/Fe] as a function of variation in $\xi_\mathrm{micro}$ color coded with \teff\, for the first 5 HF lines (R19, R18, R16, R15, and R14; top row panels) and the last 5 HF lines (R13, R12, R11, R9, and R7; bottom row panels). The The variation in [F/Fe] for +0.2 km/s change in $\xi_\mathrm{micro}$ is shown by inverted triangles and for -0.2 km/s change in $\xi_\mathrm{micro}$ by circles.}
  \label{fig:flourine_vmicdiff}%
\end{figure*}

\begin{figure*}
  \includegraphics[width=\textwidth]{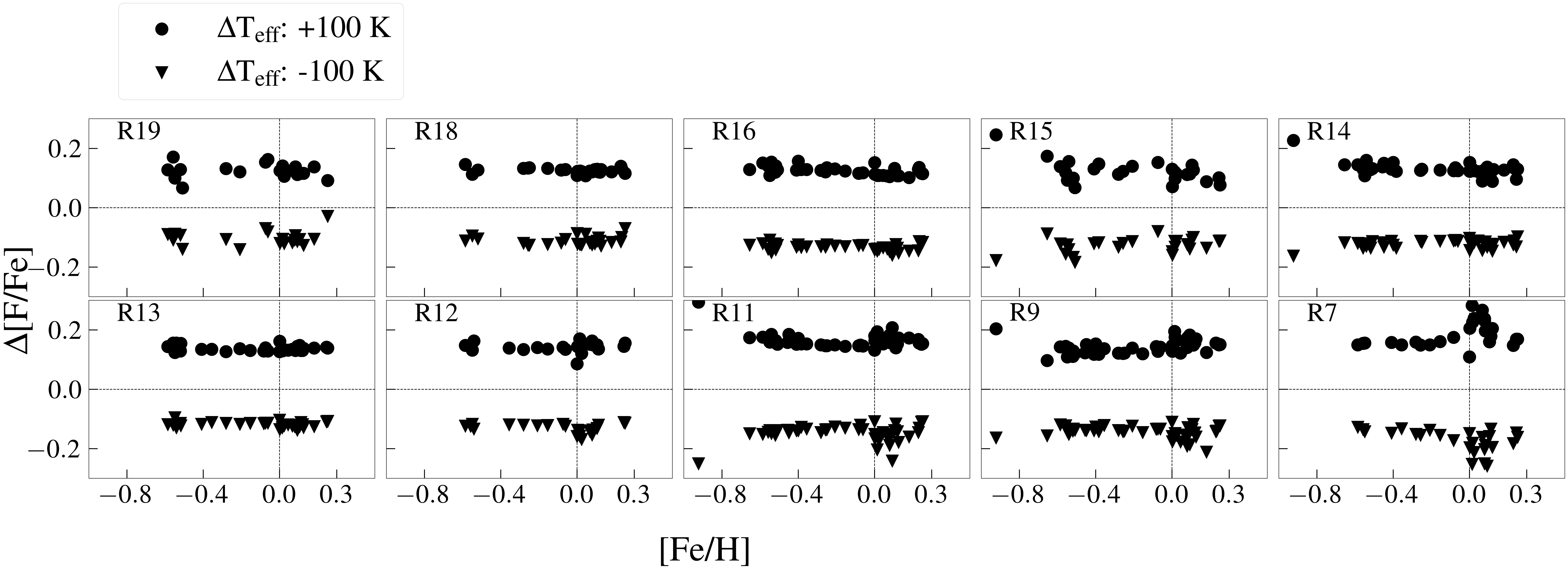}
  \caption{ Same as Figure~\ref{fig:flourine_vmicdiff} except for variation in \teff.}
  \label{fig:flourine_teffdiff}%
\end{figure*}



\section{Results}
\label{sec:results}

In this section, we want to investigate the Galactic chemical evolution trend of fluorine in the solar neighbourhood. To this end, we plot the [F/Fe] versus [Fe/H] for the fluorine abundances determined from individual HF lines in order to investigate if all of them show a similar upturn at super-solar metallicities as seen in the case of warmer stars \citep[see, e.g.,][]{Ryde:2020,Ryde:20:jaa}. We further explore the nature of the individual fluorine abundance trends with respect to \teff, \logg, \feh, and $\xi_\mathrm{micro}$, as well as their sensitivity to a change in the microturbulence, $\xi_\mathrm{micro}$, which is indicative of the degree of saturation and how in-sensitivity the line might be to the  abundance. We also explore the temperature sensitivity of the HF lines in general. The aim is to choose the best set of HF lines to determine fluorine abundances for M giants.

\subsection{Individual [F/Fe] versus [Fe/H] trends}
\label{sec:Ftrends_indiv}

In the top panels of Figure~\ref{fig:flourine_rm}, we plot the [F/Fe] versus [Fe/H] trends for the 50 solar neighborhood M giants (red circles) determined from each of the ten HF lines. The trends are quite flat with some scatter for  first four lines. The scatter then increases with increasing wavelength of the line used, especially for the four last lines. The scatter increases significantly for the metal-rich stars. 

In the bottom panel, we plot the running means of the fluorine abundances determined from each line with different colors to enable a qualitative comparison of the trends derived from the individual lines. We note that the running means plotted here have to be evaluated including the large spread, especially for the lines with the largest wavelengths. 
In this panel we also see the flat trends, especially for sub-solar metallicities. However, we also see a small fluorine enhancement at super-solar metallicities, an enhancement which increases for lines with increasing wavelengths. The lines with the largest wavelengths (the reddest ones) show the largest scatter and the largest super-solar increase. It is clear that the different HF lines show different derived fluorine abundances for the same stars at super-solar metallicity, especially from the four reddest lines at $\lambda_\mathrm{air}>23040.57$ \AA. An interesting question is then what causes these inconsistent abundances that are derived for the most metal-rich stars.  

We note that the line at 23134.76 \AA\ is heavily blended with a \co line as shown in Figure~\ref{fig:flourine_spectra}. Since other similarly weak CO lines close by are modelled well, we are confident that the fluorine abundance can be determined well from this blended line. Obviously, the blend makes it  impossible to define a mask for the HF line which is sensitive to fluorine alone. 

\subsection{Trends with stellar parameters}
\label{sec:Ftrends_parameters}

In Figure~\ref{fig:flourine_parameters}, we plot the fluorine abundances determined from each of the HF line as a function of \teff\,(top panels), \logg\,(middle panels) and $\xi_\mathrm{micro}$ (bottom panels), all color coded with the respective metallicities. 

From the figure we see that there is no significant \teff\, bias evident in our sample of the 50 stars. An exception are the four coolest stars (\teff\,$<$ 3400 K) which are all metal rich ([Fe/H] $>$ 0.0), and  show higher fluorine abundances. This should be kept in mind, since there are also metal-rich stars that are warmer which, inconsistently, do not show these high abundance values. We also note that we do not have any cool (\teff\,$<$ 3400 K), metal-poor stars in our sample. 

The abundance trends as a function of \teff\ for the 6 bluest lines (R13, R14, R15, R16, R18, and R19), ignoring the four coolest stars, show only slight slopes. Since there is a range of metallicities represented at all temperatures, apart for \teff<3400\,K, this slight trend with \teff\ will show up as a scatter in the final [F/Fe] versus \feh\ trend.  

For the 4 reddest HF lines at $\lambda_\mathrm{air} > 23000$\,\AA, (R7, R9, R11, and R12) including the commonly used line at 23358.33 \AA\ (R9), however, we see significant downward trends as a function of \teff\, with higher abundances for cooler stars and lower abundances for hotter stars.  For these lines the [F/Fe] ratio ranges from $\sim$ -0.5 to 0.5 dex for the metal-rich stars. This large spread is not seen for the other lines.     

With respect to the surface gravity, \logg, there are no significant trends in the fluorine abundances for the ten lines. There is, however,  significant scatter in the abundances from the four reddest lines at $\lambda_\mathrm{air}>23000$\AA, presumably reflecting the large abundance trends with \teff\ mentioned above. 
A similar large scatter is also evident for these four lines 
in the plot versus $\xi_\mathrm{micro}$. 


\subsection{Sensitivity to $\xi_\mathrm{micro}$}
\label{sec:Ftrends_vmic}

Next, we investigate the sensitivity of the ten HF lines to the variation in $\xi_\mathrm{micro}$. We vary $\xi_\mathrm{micro}$ by $\pm$0.2 dex, which can be considered quite a small amount for the microturbulence in general,  and determine fluorine abundances from all ten HF lines for every star. The difference in the abundances corresponding to the $\xi_\mathrm{micro}$ variation color coded in \teff\ is shown in the Figure~\ref{fig:flourine_vmicdiff}.

We see in the figure that the two lines at $\lambda_\mathrm{air}<22770$ \AA\ (R18 and R19) are insensitive to the microturbulence and therefore good for deriving an abundance from. For the sequential lines, the metal-rich stars start showing sensitivity to a change in the microturbulence. 
As we go red-wards in the series this sensitivity accelerates. There is also a clear signature that the abundances estimated for the cool (\teff\, $<$ 3500 K) stars are affected the most. 

The abundances from the HF lines at wavelengths larger than 22885 \AA\ (R7 to R14) are increasingly sensitive  to the $\xi_\mathrm{micro}$ even for lower metallicities. For the four strong lines at $\lambda>23000$ \AA, the abundance is extremely sensitive to the microturbulence, leading to $\sim$ 0.1-0.15 dex variation in [F/Fe] that increases from low to high metallicities. The strong correlation with $\xi_\mathrm{micro}$ for abundances from these four HF lines  is an indication that these lines are  strong and saturated, especially for the cool metal rich stars. 
These lines are therefore not useful for deriving an abundance from, for M giants.

We note that lower sensitivities to the $\xi_\mathrm{micro}$ have been demonstrated in earlier works by \cite{jonsson:2014b}, \cite{jonsson:2017}, and \cite{Guerco:2019b}. The stars investigated in these works are, however, warmer which means that  the lines are not saturated, hence considerably less $\xi_\mathrm{micro}$ sensitive. We find, in fact, a similar trend in which our warmer stars and more metal-poor stars show the  least $\xi_\mathrm{micro}$ sensitivity (see Figure~\ref{fig:flourine_vmicdiff}).   


\subsection{Sensitivity to the effective temperature}
\label{sec:Ftrends_teff}
\begin{figure*}
  \includegraphics[width=\textwidth]{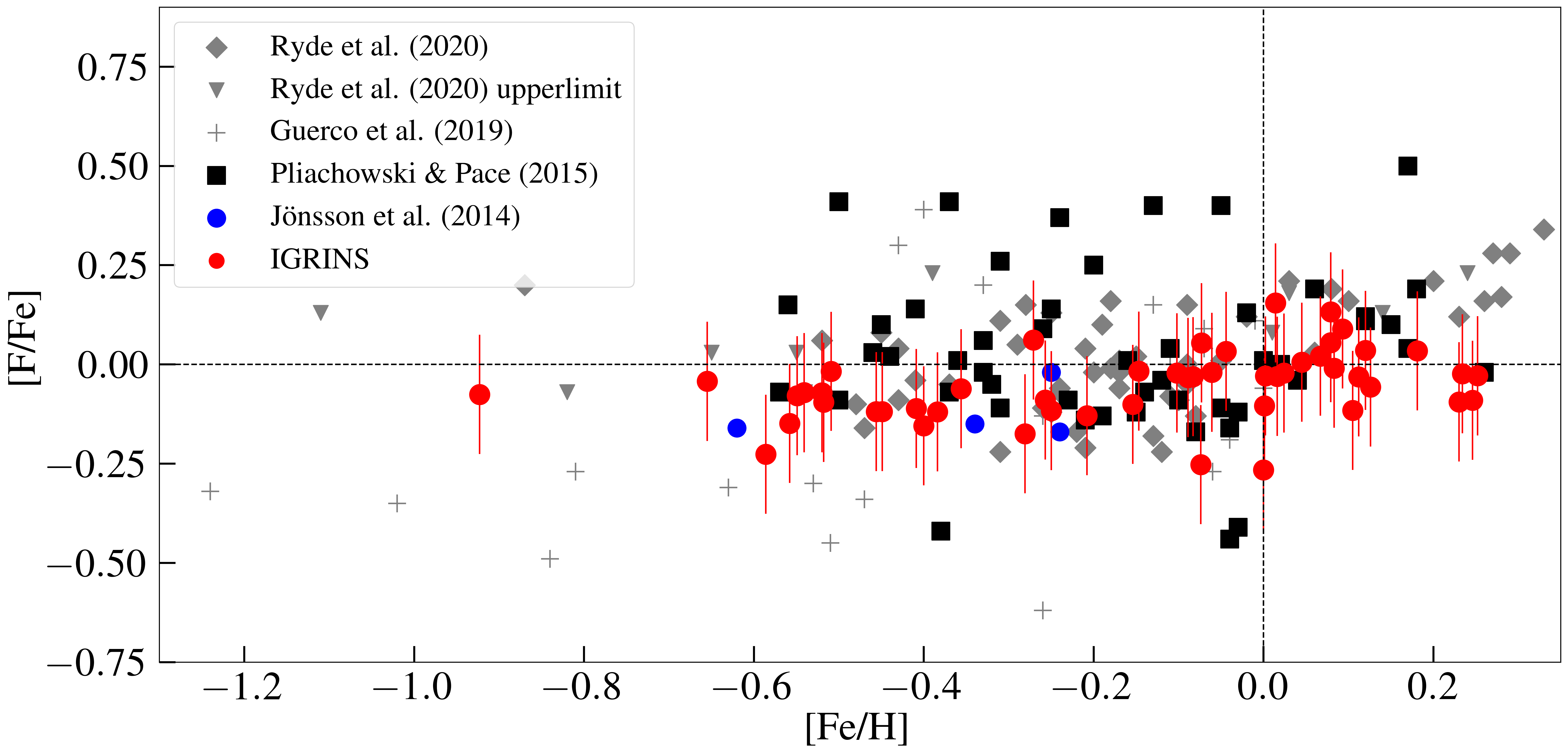}
  \caption{ Mean [Fe/Fe] versus [Fe/H] for M giants in our sample (red circles) after excluding the fluorine abundances from the four reddest and strongest HF lines at 23040.57 \AA\, (R12), 23134.76 \AA\, (R11), 23358 \AA\, (R9), and 23629.99 \AA\, (R7). Black squares denote the stars in \cite{Pilachowski:2015}, blue circles the stars in \cite{jonsson:2014b}, plus symbols the stars in \cite{Guerco:2019b}, and gray diamonds and inverted triangles the stars in \cite{Ryde:2020}.} 
  \label{fig:flourine_literature}%
\end{figure*}

Since molecular lines in general are temperature sensitive, we also investigate the sensitivity of the ten HF lines to the variation in \teff. We vary the temperature within the given uncertainties from the method deriving the effective temperatures as given in \citet{Nandakumar:2023}, i.e. $\pm100$\,K. We then determine the fluorine abundances from all ten HF lines for every star. The difference in the abundances are shown in the Figure~\ref{fig:flourine_teffdiff}.   We see, indeed, that the derived abundances vary as much as $\pm0.15$dex for a change of $\pm100$\,K. This is the largest uncertainty source in our analysis and is indicated in Figure \ref{fig:flourine_literature}.
 
\section{Discussion}
\label{sec:discussion}

In the previous section, we investigated the fluorine abundance trends from the ten HF lines as a function of stellar parameters as well as their sensitivity to $\xi_\mathrm{micro}$. Based on this, we first attempt to choose the best HF lines to determine fluorine abundances for M giants followed by the comparison of our final [F/Fe] trend as a function of [Fe/H] to those available in the literature.

\subsection{Best HF lines to determine fluorine abundances for M giants}
\label{sec:Best_HF}

From the trends with metallicity of the ten HF lines it is obvious that the lines at 23040.57, 23134.76, 23358.33, and 23629.99 \AA\ (i.e. the R7, R9, R11, and R12 lines) show a large scatter and do not follow the trends from the weaker lines. Also, from the trends with the stellar parameters, it is clear that the fluorine abundances derived from these four lines
exhibit significant unexpected and inconsistent trends with \teff, \logg, and \microturb, as well as a large scatter in these plots. The  R7, R9, R11, and R12 lines should, therefore, be avoided for abundance determinations in M giants (\teff$< 3900$\,K). A reason for these trends and scatter might be that for cool, metal-rich, and low-gravity stars, these lines are 'strong' and could, therefore, be saturated \citep[see for example][]{gray}. This is actually indeed reflected in the large spread in the microturbulence trends as seen in Figure~\ref{fig:flourine_vmicdiff}. 
The abundances from these lines demonstrate strong correlations with $\xi_\mathrm{micro}$, especially for cool, metal-rich stars. 

We also find that the abundances from the three most blue-ward lines at 22699.49, 22714.59, and  22778.25 \AA\, (R19, R18, and R16) are the least susceptible to trends with \teff, \logg\, and $\xi_\mathrm{micro}$. These should therefore, primarily, be used for abundance determinations for M giants. However, since these lines are increasingly weak, the slightly stronger lines at 22826.86, 22886.73, and 22957.94 \AA\ (R15, R14, and R13)  could also be used, but with caution. The more red lines should always be avoided for M giants. 

We also find that the strengths of the vibration-rotational lines of the HF molecule are very sensitive to uncertainties in the effective temperatures of the stars. This is not unexpected for molecules \citep[OH is also very temperature sensitive, see e.g.][]{Nandakumar:2023}, which are sensitive to the temperatures in the line-forming regions. This could either be a result of uncertainties in the stellar parameters or in the detailed temperature structure itself of the model atmospheres. An intrinsic problem with the molecular lines of HF is that the stellar temperatures are needed to be known with a high accuracy, which is difficult to achieve. Thus, the slight slopes in the abundance versus temperature plots might arise from a small systematic uncertainty in the \teff determination of the stars from \citet{Nandakumar:2023}. 




We note that there is heavy \co\, blend in the line at 23134.76 \AA\, (R11, see Figure~\ref{fig:flourine_spectra}). This blend can be taken care of provided that the similar CO lines close by in wavelength are modelled well. This ensures that the blend is properly modelled together with the HF line \citep[this was successfully done for other elements in e.g.][]{nandakumar:22,montelius:22}.




\subsection{[F/Fe] versus \feh: comparison to literature}
\label{sec:Best_HF}

Due to the reasons elaborated in the previous section, we neglect the abundances from the four problematic lines (R7, R9, R11, and R12) for all stars in our sample while plotting the final [F/Fe] versus \feh\, trend. Further, we only use the abundances from the three bluest HF lines (R16, R18,
and R19) for cool and metal rich stars, i.e. \teff\, $<$ 3500 K and \feh\,$>$ 0.0 dex. For all other stars, we plot the mean [F/Fe] estimated from the six HF lines selected based on their quality (see Table~\ref{table:lines}). We show the final [F/Fe] versus \feh\, for the 50 stars in our sample in the Figure~\ref{fig:flourine_literature}. As the comparison sample, we plot the fluorine abundances determined for stars in \cite{jonsson:2014b} (blue circles), \cite{Pilachowski:2015} (black squares), \cite{Guerco:2019b} (grey plus symbols) and \cite{Ryde:2020} (grey diamonds and inverted triangles).

\cite{jonsson:2014b} determined fluorine abundances from the R9 line using spectrum synthesis with SME for four bright, nearby giants (including Arcturus) by analysing their spectra observed with the
Fourier Transform Spectrometer (FTS) mounted on the Kitt Peak National Observatory Mayall 4 m reflector. The \teff\, values were determined using angular diameter measurements taken from \cite{Mozurkewich:2003}, and the surface gravity, \logg, was based on the stellar radius, the parallax, and fits to evolutionary tracks. The metallicities were adopted from different spectroscopic archives and literature sources. All four stars have sub-solar metallicities, and apart from Arcturus with \teff\, of 4226 K the stars lie in the effective temperature range of 3700 to 3900 K. 

 \cite{Pilachowski:2015} determined the fluorine abundances also from the R9 line for nearly 80 G and K stars in the Galactic thin disk from spectra obtained with the Phoenix IR spectrometer on the 2.1 m telescope at Kitt Peak. They carried out the analysis using both
spectral synthesis (with MOOG) and an equivalent width analysis (for weak lines). The stellar parameters for these stars have been adopted from various sources in the literature, with \teff\, in the range of 3800 K to 4800 K and metallicities in the range -0.6 $<$ \feh\, $<$ 0.3 dex. They estimated an 
average fluorine abundance of [F/Fe] = +0.23 $\pm$ 0.03 in the thin disk with a large scatter. In Figure~\ref{fig:flourine_literature}, we plot [F/Fe] for a subset of 52 stars, omitting the stars with upper limits. 

\citet{Guerco:2019b} estimated the fluorine abundances for a sample of Milky Way red giants from up to five HF lines (R16, R15, R14, R13, and R9) in K band spectra obtained by observations using three infrared spectrographs: NOAO Phoenix \citep{phoenix}, iSHELL
spectrograph \citep{ishell}, and the Fourier Transform Spectrometer (FTS) archive \citep{fts}. The \teff\, values, determined from photometric calibration, optical spectra, or adopted from APOGEE as well as various other literature sources, range from 3400 K to 4900 K. The metallicities of the analysed stars are limited to metal-poor to solar metallicities (\feh$ < 0.0$) as shown in Figure~\ref{fig:flourine_literature}. 

\cite{Ryde:2020} analysed high resolution K-band spectra of 61 Milky Way K-giants obtained using the IGRINS and Phoenix spectrographs, and determined the fluorine abundances from the HF line at 23358.33 \AA\ (R9). They did not use other HF lines since the stars in their sample are warmer than the stars we have analysed here (4000 K to 4600 K), resulting in weaker HF lines at shorter wavelengths. They also provided an upper limit of the fluorine abundances for several stars with very weak 23358.33 \AA\, lines (mostly hot, metal poor, and high \logg\, stars). A majority of the metal-rich stars in their study have the warmest \teff\, in their sample (\teff\, $>$ 4300 K). Thus, for these stars, the HF line  is much weaker than for M giants in our sample and can therefore not be expected to be saturated like the case for cool M giants. Hence, the fluorine abundance determined from this line should be reliable. The stellar parameters in the \cite{Ryde:2020} sample were determined from careful analysis from high-resolution optical spectra \citep{jonsson:17}. The stars in their sample cover a broader metallicity range of $-1.2 <$ \feh\, $<0.4$.

The fluorine abundances determined for the four stars in \cite{jonsson:2014b} are consistent with our values in the same metallicity range. Our measurements are also consistent with [F/Fe] values measured by \cite{Pilachowski:2015}, although these show a larger scatter. Our final fluorine abundance trend agrees with the flat trend in \cite{Ryde:2020} in the subsolar metallicity range, i.e. for $-1.0 <$ \feh\ $< 0.0$. The scatter in the trend of \cite{Ryde:2020} is larger (several stars with supersolar [F/Fe]), which is not the case with our trend. The sample from \cite{Guerco:2019b} also exhibit a similarly flat trend for \feh$>$ -0.4, but decreases by $\sim$ 0.1 - 0.2 dex for more metal-poor stars. This is a larger decrease than we see in that metallicity range. 
At super-solar metallicities, there is a clear increase in [F/Fe] for stars from \cite{Ryde:2020}, while we find a slight enhancement ([F/Fe] $\sim$ 0.1 dex) soon after solar metallicity. This could thus confirm a slight enhancement compared to more metal-poor stars. However, for stars more metal rich than \feh=0.15, we can not confirm the steady and clear increase in [F/Fe] as seen in \citet{Ryde:2020}. The stars in our sample that have \feh $>0.2$ are warmer than 3550\,K and are therefore not affected that much by uncertainties in the microturbulence. Our derived fluorine abundances are therefore reliable.
Our sample lack stars with \feh\, higher than 0.25 dex to compare with similar stars in \cite{Ryde:2020}. All HF lines are however temperature sensitive and it has to be investigated whether the difference at \feh $>0.2$ in these studies is due to uncertainties in the temperatures or whether it could be an unknown blend in the R9 line that mostly affects the warmer stars.  

To conclude, we find a flat [F/Fe] versus \feh\ trend with no significant upturn at super solar metallicities, but a hint of a slightly higher level of [F/Fe] for stars with  \feh\ $>0$. The spread in our trend, which is smaller that that of \citet{Ryde:2020}, is presumably caused by the temperature sensitivity of the HF line. With the level of accuracy we have for the stellar parameters such a spread is indeed expected.

\section{Conclusions}
\label{sec:conclusion}

We have carried out a detailed line-by-line abundance analysis of ten molecular HF lines in the IGRINS spectra of 50 Milky Way M giants (3300 K $<$ \teff\, $<$ 3800 K) in order to investigate the nature of the fluorine abundance trend as a function of metallicity. Among the 50 stars, 44 stars have been observed within the programs GS-2020B-Q-305 and GS-2021A-Q302, while six have been extracted from the IGRINS spectral library. The stellar parameters for these stars have been determined based on the method using selected OH, CN, CO and Fe lines in the H band spectra described in \cite{Nandakumar:2023}. 

Based on our investigations,  we find that it is important to choose the lines wisely after detailed quality checks. We conclude that the R19, R18, and R16 lines (22699.49, 22714.59, and  22778.25 \AA) of HF should primarily be used for abundance determinations for M giants, and can be used even for cool, metal-rich stars. These are the least susceptible to trends with \teff, \logg\, and $\xi_\mathrm{micro}$. 
However, since these lines are increasingly weak, the slightly stronger R15, R14, and R13  lines at 22826.86, 22886.73, and 22957.94 \AA\  could also be used, especially for lower metallicities. 
The strongest HF lines  (R7, R9, R11, and R12) have significant trends with the stellar parameters, as well as a high sensitivity to variations in $\xi_\mathrm{micro}$, especially for coolest and most metal-rich stars.
These lines should, therefore, be avoided for abundance determinations in M giants (\teff < 3900 K). 


Apart from the large trends of the derived abundances with the effective temperature of the stars for the four strongest HF lines, we also find a slight systematic trend with \teff\ for the six weaker lines. A reasonable explanation might be that this is caused by the large sensitivity of the molecular lines to the effective temperature and/or uncertainties in  the  temperature structure of the model atmosphere. These trends lead to an increased spread in the [F/Fe] versus [Fe/H] plots.
This sensitivity is an intrinsic problem with lines from the HF molecule. 
One way to mitigate this problem would be to choose a sample of stars with a narrow range in \teff. 

For cool (\teff\, $<$ 3500 K), metal-rich stars, the HF lines (except R16, R18, and R19) show a higher sensitivity of the derived fluorine abundances to variations in $\xi_\mathrm{micro}$. This indicates that these lines might be strong or saturated, making their abundances uncertain. Thus we neglected the fluorine abundances from the fours strongest lines (R7, R9, R11, and R12) for all stars and used only those derived from R16, R18, and R19 for the coolest and most metal-rich stars when estimating the final mean fluorine abundance trend as a function metallicity. 

From this trend, we confirm the flat trend of \citet{Ryde:2020} and \cite{Guerco:2019b} for stars in the metallicity range of -1.0 $<$ \feh\,$<$ 0.0  and -0.6 $<$ \feh\,$<$ 0.0  respectively. We also find a slight enhancement at supersolar metallicities (0<\feh<0.15), that might be consistent with the general trend of \citet{Ryde:2020}. 
However, using our limited sample of stars at super-solar metallicities, we cannot confirm the upward trend seen at \feh\,$>$ 0.15 in \cite{Ryde:2020}. We find lower fluorine abundances at these higher metallicities. Hence we need more observations of many more M giants especially at high metallicities with their spectra from a capable high resolution instrument like IGRINS to disentangle the correct nature of the fluorine trend.

\begin{acknowledgements}
 Henrik Jönsson is thanked for enlightening discussions. G.N.\ acknowledges the support from the Wenner-Gren Foundations and the Royal Physiographic Society in Lund through the Stiftelsen Walter Gyllenbergs fond. N.R.\ acknowledge support from the Royal Physiographic Society in Lund through the Stiftelsen Walter Gyllenbergs fond and Märta och Erik Holmbergs donation.  This work used The Immersion Grating Infrared Spectrometer (IGRINS) that was developed under a collaboration between the University of Texas at Austin and the Korea Astronomy and Space Science Institute (KASI) with the financial support of the US National Science Foundation under grants AST-1229522, AST-1702267 and AST-1908892, McDonald Observatory of the University of Texas at Austin, the Korean GMT Project of KASI, the Mt. Cuba Astronomical Foundation and Gemini Observatory.
This work is based on observations obtained at the international Gemini Observatory, a program of NSF’s NOIRLab, which is managed by the Association of Universities for Research in Astronomy (AURA) under a cooperative agreement with the National Science Foundation on behalf of the Gemini Observatory partnership: the National Science Foundation (United States), National Research Council (Canada), Agencia Nacional de Investigaci\'{o}n y Desarrollo (Chile), Ministerio de Ciencia, Tecnolog\'{i}a e Innovaci\'{o}n (Argentina), Minist\'{e}rio da Ci\^{e}ncia, Tecnologia, Inova\c{c}\~{o}es e Comunica\c{c}\~{o}es (Brazil), and Korea Astronomy and Space Science Institute (Republic of Korea).
The following software and programming languages made this
research possible: TOPCAT (version 4.6; \citealt{topcat}); Python (version 3.8) and its packages ASTROPY (version 5.0; \citealt{astropy}), SCIPY \citep{scipy}, MATPLOTLIB \citep{matplotlib} and NUMPY \citep{numpy}.
\end{acknowledgements}

%
%


\bibliographystyle{aa}
\bibliography{references} 




\end{document}